# Elastic relaxation during 2D epitaxial growth: a study of in-plane lattice spacing oscillation

P. Müller[1], P. Turban[2], L. Lapena[1], S. Andrieu[2]

1) *Centre de Recherche sur les Mécanismes de la Croissance Cristalline*[1], CRMC2– CNRS, Campus de Luminy, Case 913, 13288 Marseille CEDEX 9, France

2) *Laboratoire de Physique des Matériaux*, UMR7556, Univ. H. Poincaré, F-54506 Vandoeuvre

## Abstract

The purpose of this paper is to report some new experimental and theoretical results about the analysis of in-plane lattice spacing oscillations during two-dimensional (2D) homo and hetero epitaxial growth. The physical origin of these oscillations comes from the finite size of the strained islands. The 2D islands may thus relax by their edges, leading to in-plane lattice spacing oscillations during the birth and spread of these islands. On the one hand, we formulate the problem of elastic relaxation of a coherent 2D epitaxial deposits by using the concept of point forces and demonstrate that the mean deformation in the islands exhibits an oscillatory behaviour. On the other hand, we calculate the intensity diffracted by such coherently deposited 2D islands by using a mean model of a pile-up of weakly deformed layers. The amplitude of in-plane lattice spacing oscillations is found to depend linearly on the misfit and roughly linearly on the nucleation density. We show that the nucleation density may be approximated from the full-width at half maximum of the diffracted rods at half coverages. The predicted dependence of the in-plane lattice spacing oscillations amplitude with the nucleation density is thus experimentally verified on V/Fe(001), Mn/Fe(001), Ni/Fe(001), Co/Cu(001) and V/V(001).

---

[1] *associated to the Universities of Aix-Marseille II and III*

*Email: muller@crmc2.univ-mrs.fr*

**Introduction**

Considerable progress in growing thin films of metals, oxides and semi-conductors have been performed during the past 30 years, due to both the richness in new fundamental phenomena and the large potential of applications. Thin films are often grown in a polycristalline state, but single-crystalline films are sometimes needed for microelectronics for instance. On a fundamental view-point, epitaxial systems are also needed since they are model systems to study new physical and chemical phenomena due to the interfaces, the reduced size of the films and the presence of the surface. The general problem of epitaxial growth was thus largely investigated on films grown by Molecular Beam Epitaxy (MBE). Two general processes may be distinguished in the epitaxial 2D growth of a material A on a substrate B. A first regime occurs at the beginning of the growth, where A adopts the in-plane lattice spacing of B. This regime is usually called the pseudomorphic regime, corresponding to the accumulation of elastic energy when the thickness is increased. When this elastic energy becomes higher than the energy necessary to create dislocations (and if the kinetics conditions allow to create them), the material A relaxes towards its regular parameter so that its in-plane parameter varies as a function of the film thickness. This second regime is usually called the relaxed regime associated to plastic relaxation (for a review see [1]). During the last 20 years, a lot of work have been devoted to the understanding of the phenomena which occur during such plastic relaxation. Because strained crystals may have different structural or physical properties than strain-free crystal, the pseudomorphic regime was also thoroughly studied (for a review see [2]). In this case, owing to the pseudomorphy, no in-plane lattice spacing variation was expected before the plastic relaxation occurs. However, studying the epitaxial growth of $In_xGa_{1-x}As$ on lattice mismatched GaAs(001) where in the 2D pseudomorphic regime growth takes place by birth and spread of 2D islands, Massies and Grandjean [3] discovered, by using RHEED, an associated in-plane lattice spacing oscillatory behaviour. Such in-plane oscillations have been confirmed by Eymery et al [4] Fassbender et al [5], Hartmann et al [6] and P.Turban et al [7] on many other systems.

The first qualitative explanation was given by Massies and Grandjean [3]: In contrast to a continuous layer, 2D strained islands may elastically relax by their free edges. As a consequence, during lateral growth of 2D islands towards a continuous layer, the mean in-plane lattice spacing oscillates. For the authors [3] the surface parameter deviation would be maximum (minimum) for half (complete) coverage as the step density should be. Another more quantitative approach [8], but based on a phenomenological approach of the electron diffraction, allowed us to predict that in-plane lattice spacing oscillation might occur even for homoepitaxy (that means in absence of natural misfit $m_o$) and/or at constant free edge density. Recent experimental data confirm these points [7]. However, a quantitative treatment of the experimental data and especially a comparison of the effect from one system to another was difficult essentially because the islands size (or the nucleation density) have to be known.

The purpose of this paper is to go further into the description of in-plane lattice spacing oscillations, both on the experimental and theoretical view-points. More precisely we want to give an answer to the following questions: (i) What are the physical origins of the oscillations of position and width of the RHEED streaks ? (ii) Why homoepitaxial systems (that means with zero misfit) also exhibits oscillations (iii) What are the dependence of these oscillations with misfit, nucleation density, island size …? For this purpose in section I we reformulate the problem of elastic relaxation of coherent 2D epitaxial deposits in a more complete form than in [8]. For a good understanding of the underlying physics we will prefer an analytical formulation easier to discuss than simulation results. Thus we only consider the case of epitaxial deposit that consists in periodic 1D ribbon (one monolayer thick) where lateral growth takes place at constant steps density (excepted at coalescence). We believe that such model allows us to capture the essential physics. In section II we compare the experimental data to theoretical one. First, we demonstrate experimentally that the full-width at half maximum (FWHM) is actually connected with the nucleation density. Second, we show that the amplitude of the detected relaxation effect actually depends on the nucleation density, or in other words, on the 2D islands size, in agreement with the theory.

**I/ Model**

In order to compare experimental data on RHEED rod-spacing oscillations with calculations, we proceed in 3 steps: **(1)** we show how the finite lateral size of the 2D islands plays on the misfit definition. Moreover, we show that an elastic misfit may be defined even for homoepitaxy since a small 2D island has a different in-plane lattice spacing than in the bulk (section I1). **(2)** by using the concept of point forces, we formulate the problem of elastic relaxation of coherent 2D epitaxial deposits. The corresponding mean deformation in the islands actually leads to an oscillatory behaviour with coverage (section I2). **(3)** by using a mean model of a weakly deformed layers pile-up, we calculate the intensity diffracted by such coherently deposited 2D islands and give analytical expressions of the rod-to-rod distance oscillation of the corresponding RHEED pattern (section I3).

### I1/ Epitaxial misfit

The natural misfit $m_o$ of two infinite cubic phases A (crystallographic parameter $a_o$) and B (parameter $b_o$) is defined for parallel axis epitaxies as:

$$m_o = (b_o - a_o)/a_o \qquad (1)$$

Thus the parameters $a_o$ and $b_o$ are linked by $b_o = a_o(1+m_o)$. However because of its broken bonds, a small piece of matter cut in the infinite phase A may relax to reach its own mechanical equilibrium. In other words the crystallographic parameter $a$ of a finite-size crystal (volume $V = h l_1 l_2$) cut in the infinite phase may differ from $a_o$. Let us note $a = a_o [1 + \varepsilon(h, l_1, l_2)]$ where $\varepsilon(h, l_1, l_2)$ is the size dependent strain with respect to the infinite "mother-phase".

Obviously the smaller the crystal is, the greater $\varepsilon(h, l_1, l_2)$ as has been observed on many systems (see for example [9,10]). Therefore for a finite size crystal A grown onto a semi-infinite substrate B, the active misfit, $m$, has to be distinguished from the natural misfit $m_0$ [8] as :

$$m = (b_0 - a)/a \approx m_o - \varepsilon(h, l_1, l_2) \qquad (2)$$

from the natural misfit $m_o$. *It should be noted that the active misfit may exist even in the case of homoepitaxy ($m_o=0$). This means that a crystal A of finite size has to be strained by the quantity $m = -\varepsilon(h, l_1, l_2)$ to be accommodated on its own substrate A.* Obviously since $\varepsilon(h, l_1, l_2)$ is size dependent this effect only exists for islands of nanometric dimensions.

For small but macroscopic 3D crystals where surface energy and surface stress quantities make sense, $\varepsilon(h, l_1, l_2)$ only depends on the surface stress of the lateral ($s_A'$) and basal ($s_A$) faces of the crystal. For instance, it reads for a quadratic crystal $V = h l^2$ [8]

$$\varepsilon(h,l) = -\frac{1-\nu_A}{E_A}\left(\frac{2s_A}{h} + \frac{2s'_A}{l}\frac{1-3\nu_A}{1-\nu_A}\right) \quad (3)$$

where $E_A$ and $\nu_A$ are the Young modulus and Poisson ratio of A respectively. Thus a small crystal of A having positive surface stress (as it is generally the case for clean surfaces (see [9-11])) has a smaller lattice spacing than the infinite phase. For nanometric size the difference amounts 1%, whereas for more than millimetric crystal there is no more reason to discriminate m from $m_o$.

Since in-plane lattice oscillation occurs in the 2D growth mode, we deal in the following with 2D islands (h=a) of finite lateral size $l$ where active misfit m differs from natural misfit $m_o$. Therefore:

$$a_{2D} = a_o(1 + \varepsilon(a,l)) \quad (4)$$

However the concept of surface stress (that is a surface excess quantity [12]) for nano-crystals of a few atoms is quite questionable so that equation (3) cannot be extrapolated to define the crystallographic parameter $a_{2D}$ of a 2D crystal of smallest size. Nevertheless we show in appendix I that the finite size effect may always be written as:

$$\varepsilon(a,l) = C_o + C_1/l \quad (5)$$

where $C_o$ and $C_1$ can be estimated in a microscopic (atomistic model). It should be noted that excepted for very small lateral sizes (a few atoms), the term $C_1/l$ is only a weak correction to the strain. In the following (excepted when clearly mentioned in the text), we take $\varepsilon(a,l) = C_o$ which means that the misfit is independent on the lateral size of the island.

## I2/ Equilibrium strain in deposited islands
### I21/ Displacements and deformation fields:

Owing to their active misfit (even in the homoepitaxial case), 2D islands have to be strained to be put in coherence on their substrate. Obviously these islands may elastically relax because of their free edges. In this section we calculate the equilibrium strain in the deposited islands for a given active misfit. For this purpose, we only consider the case of coherent epitaxy which means that the island-substrate interface is assumed to be coherent and to remain coherent during the elastic relaxation. In other words, a continuity of the displacement across the island-substrate interface is assumed (no dislocations in the islands). During the elastic relaxation, the relaxing islands thus drag the atoms at the contact area, producing a strain field in the substrate. The elastic interaction between the deposit and substrate is modelled by point forces due to active misfit.

For the sake of simplicity, we only treat the case of infinite ribbons of material A deposited onto a mismatched substrate B (see figure 1). The elastic interaction between such misfitted 2D islands and the underlying coherent substrate is modelled by a distribution of lines of point forces located at the edges of the deposited islands on the substrate surface. More precisely, the forces exerted on the substrate by a set of periodic (L apart) 1D stressed ribbons (width $l$) can be written as [13,14]:

$$F_x = \frac{\partial}{\partial x}\left(\sigma_{xx}^A a\right) = \sigma_{xx}^A \, a \sum_{n=-\infty}^{\infty}\left[\delta(x+nL+l/2) - \delta(x+nL-l/2)\right] \tag{6}$$

where $\sigma_{xx}^A = E_A m/(1-\nu_A^2)$ is the in-plane stress before relaxation, $\delta(x)$ the Dirac function and $a$ the atomic thickness of the deposited island A. The in-plane stress distribution $\sigma_{xx}^B$ induced in the underlying substrate B is connected to the force distribution $\vec{F}(F_x,0,0)$ applied on the surface of B by way of the 2D Green tensor [15,16]:

$$\sigma_{xx}^B(x, z=0) = -\frac{2}{\pi} \int_{-\infty}^{\infty} \frac{F_x}{x-u} du \tag{7}$$

By using this approach, the local atomic displacement as well as the mean deformation over the surface can be calculated. In the following, we first calculate the local displacements needed to calculate the diffracted response. Moreover, the equations of the mean deformation are also needed in order to understand the interrelation between the strained islands and the underneath layers.

Putting (6) in (7) in 1D geometry (more precisely in plane strain conditions $\varepsilon_{yy}(x,z)=0$ and $\sigma_{zz}(x,z=0)=0$ for mechanical equilibrium) and since $\varepsilon_i = \sigma_i(1-\nu_i^2)/E_i$, the in-plane deformation at the surface of B (z=0) reads:

$$\varepsilon_{xx}^B(x, z=0) = \frac{2}{\pi} Kma \frac{4}{l} \sum_{n=0}^{\infty}\left\{\left[\left(\frac{x+nL}{l/2}\right)^2 - 1\right]^{-1} + \left[\left(\frac{x-nL}{l/2}\right)^2 - 1\right]^{-1}\right\} \tag{8}$$

where $K = \dfrac{(1-\nu_A^2)/E_A}{(1-\nu_B^2)/E_B}$ is the relative rigidity of B with respect to A.

The corresponding in-plane displacement (at the surface z=0) $u_x^B(x) = \int_0^x \varepsilon_{xx}^B(x, z=0)dx$ thus reads:

$$u_x^B(x) = u_x^0(x) + u_x^{Int}(x) \qquad (9)$$

where the first term $u_x^0(x)$ is the displacement field in absence of elastic interaction between the islands (or in other words when there is a single island), and the second term $u_x^{Int}(x)$ the part due to the elastic interactions with the other islands. All theses terms are given in table I.

It should be noted that there is some invariant points $u_x^B(0) = u_x^B(L/2) = 0$ [2] due to the symmetry (fig.1). Furthermore there is the asymptotic behaviour $\lim_{l \to L}[u_x^B(x)] = 0$ which means that the film becomes pseudomorphous to its substrate at coalescence. This important behaviour originates in the elastic interaction in between the islands via the substrate deformation since $\lim_{l \to L}[u_x^B(x) = u_x^0(x)] \neq 0$ when the elastic interactions are neglected. Also interesting for the following is to calculate the mean deformations at the surface z=0 $\langle \varepsilon \rangle = \frac{2}{l} \int_0^{l/2-a} \varepsilon_{xx}^B(x,0)dx$ and $\langle \varepsilon \rangle_{out} = \frac{2}{L-l} \int_{l/2+a}^{L/2} \varepsilon_{xx}^B(x,0)dx$ where $\langle \varepsilon \rangle$ is the deformation in B areas covered by the A islands, and $\langle \varepsilon \rangle_{out}$ the deformation in B uncovered areas. Moreover, a cut off distance *a* is introduced to avoid local divergences. This also allows us to recover that the mean deformation $\langle \varepsilon \rangle$ at the surface z=0 of B under the island A is equal to the mean deformation in the island with respect to the substrate B (epitaxial coherence). Like for the local displacement, these quantities read :

$$\langle \varepsilon \rangle = \langle \varepsilon \rangle^0 + \langle \varepsilon \rangle^{Int}, \qquad \langle \varepsilon \rangle_{out} = \langle \varepsilon \rangle_{out}^0 + \langle \varepsilon \rangle_{out}^{Int} \qquad (10)$$

where $\langle \varepsilon \rangle_i^0$ is the mean deformations in absence of elastic interaction (subscript 0) and $\langle \varepsilon \rangle_i^{Int}$ the interaction contribution (subscript Int). Their analytical descriptions are given in table II.

**I22/ Discussion**

First, it seems necessary to verify that the calculated atomic displacements lead to appropriate predictions according to the hypotheses. In figure 2 the displacement field $u_x^B(x)$ is plotted as the sum of two contributions, one directly due to the islands $u_x^0(x)$ and the other due to the elastic interactions between islands $u_x^{Int}(x)$. More precisely we plot the normalised displacement $u_x^i(x)/(Kma/\pi)$ as a function of $0 < x < L$ for various values of surface coverage $\theta = l/L$. For our purpose we take L=100 and θ=0.1 (fig.2a) , 0.5 (fig. 2b) and 0.99 (fig. 2c). It can be easily seen in figure 2 that **(i)** the displacement field $u_x^B(x)$ vanishes at x=0 and x=L/2 (50 in fig 2) when the elastic interactions are taken into account, **(ii)** an isolated island having a positive (negative) misfit induces a compression (dilatation), $u_x^0(x)$ <0 (>0) of the underlying substrate, **(iii)** the elastic interaction opposes to these behaviour since in all the cases $u_x^0(x)$ and $u_x^{Int}(x)$ have opposite sign. This can be easily understood since the substrate strains created by each island in the substrate overlap near the coalescence, creating a back-stress effect (see [8]). Obviously the closer the islands are, the greater the back stress effect. **(iv)** At the limit of a continuous layer ( see fig. 2c where θ=0.99) there is no more substrate deformation since the continuous layer becomes pseudomorphic to its substrate.

---

[2] It is not the case in [8] where displacement fields under and outside the islands have not been obtained within the same approximation.

Second, we are interested in the general predictions using this elastic model. In figure 3a is plotted the normalised displacement $u_x^B/(Kma/\pi)$ calculated at the island edge $x = l/2$ as a function of θ for L=100 (more precisely we calculate $u_x^B(l/2 - a) = \frac{l}{2}\langle\varepsilon\rangle_{in}$ to avoid the local discontinuity at $x = l/2$ ). In addition to the above-mentioned remarks (i) to (iv) we may see that when elastic interactions are properly considered, the edge displacement passes through an extremum (a minimum for positive misfit, a maximum for negative misfit) at $du_x^B/d\theta = 0$ for θ=1/2 according to equations of table II. Once again such extremum does not exist when elastic interactions are neglected. According to elastic interactions in between islands via the underlying substrate, the edge displacement is thus found maximum at half coverage even when the growth takes place at constant step density. In figure 3b the normalised mean deformation in the island $\langle\varepsilon\rangle/(Kma/\pi)$ is also reported as a function of coverage θ for various values of L. It is easy to see that the mean deformation also exhibits an oscillatory behaviour. For a given coverage θ the greater the L value is, the smaller the value of the mean deformation since at the limit $l = L\theta \to \infty$, the film becomes pseudomorphous so that $\langle\varepsilon\rangle \to 0$. Obviously $\langle\varepsilon\rangle$ and $\langle\varepsilon\rangle_{out}$ behave in a complementary way since the total mean deformation $\langle\varepsilon\rangle_L$ over a period L of a complete substrate layer reads:

$$\langle\varepsilon\rangle_L = \theta\langle\varepsilon\rangle + (1-\theta)\langle\varepsilon\rangle_{out} = 0 \qquad (11)$$

*In other words the mean deformation of the top substrate layer is zero whereas the mean deformation (with respect to the substrate parameter b) of a coherently supported island (l < L) is not.* The deformations in the underlying layers of the substrate have to be calculated by using the z-dependent Green tensor [15,16]. Nevertheless at each level z, $\langle\varepsilon(z)\rangle$ and $\langle\varepsilon(z)\rangle_{out}$ behave in a complementary way so that the mean deformation $\langle\varepsilon\rangle_L$ of each substrate layer is in fact equal to zero.

## I3/ Diffracted intensity

Our purpose in this section is to determine the electron intensity scattered by such deposited islands which deforms the underlying substrate during their elastic relaxation. In order to capture the essential physic with analytical expressions **(i)** *we model* the system by a pile-up of weakly deformed layers described from an average lattice (see figure 4), **(ii)** we use the kinematical scattering theory which is valid for the determination of reflection positions even for electron diffraction but does not generally lead to accurate intensity calculations [17], **(iii)** Nevertheless in order to restore suitable usual RHEED oscillations intensity, we will correct the classical kinematical theory by including refractive corrections [17] and using the top-layer interference model [18]. More precisely in section I31 we show that the substrate layers behave (for diffraction) as non deformed layers because of their zero mean deformation, whereas the growing layer behaves as a mean deformed layer. In section I32 we therefore consider the diffraction properties of such a pile-up of layers (substrate + growing layer). At last in section I33 we comment the oscillation of full-width at half maximum of the RHEED rods.

### I31/ Diffraction by a weakly deformed layer

The calculation of the scattered intensity is reported in appendix II. We consider a 2D imperfect layer whose in-plane lattice spacing is referred to the one of a perfect crystal as $\vec{x}_n = n\vec{a} + \vec{u}_x^B(x=na)$. $\vec{u}_x^B(x=na)$ is the in-plane displacement vector (see (9)) induced in the surface of the substrate B by the deposited islands. This intensity thus reads (in appendix II $\vec{u}_x^B(x = na)$ is noted $\vec{u}_n$):

$$I = f^2 \sum_p \exp(2i\pi\vec{s}.\vec{x}_p) + \sum_N \sum_p f^2 4\pi^2 q^2 \left(\frac{2Km}{\pi}\right)^2 \frac{1}{2}\left(\frac{\sin(\pi N\theta)}{N}\right)^2 \left[\exp(2i\pi(\vec{s}+N\vec{k}).\vec{x}_p) + \exp(2i\pi(\vec{s}-N\vec{k}).\vec{x}_p)\right]$$

It is the sum of 2N+1 terms. The first term is the regular diffraction of the average lattice. The N other terms correspond to satellites whose positions are deduced from the node of the average lattice by the translation ±Nk with a structure factor equal to $f^2 4\pi^2 q^2 \left(\frac{2Km}{\pi}\right)^2 \frac{1}{2}\left(\frac{\sin(\pi N\theta)}{N}\right)^2$. The intensity ratio of the $N^{th}$ satellite to that of the average node thus reads $I/I_o = \frac{q^2 K^2 m^2}{2}\left(\frac{\sin(\pi N\theta)}{N}\right)^2$. It is misfit square dependent and increases with the diffraction order as $q^2$. The diffracted spectrum exhibits no satellite around the centre (q=0) but exhibits very weak satellites for q=±1 order (which is the single order generally recorded with the RHEED experiment). For usual misfit (m= $10^{-2}$) the more intense satellite N=1 (located at ±1/L with respect to the peak q=±1) scale as $I/I_o = 10^{-4}$. However these weak satellites cannot be observed in usual RHEED experiments[3] because of the poor resolution of usual RHEED detectors. *In the following, we consequently consider that the RHEED intensity scattered by a weakly deformed layer is the one of the average lattice. Indeed, as the mean deformation of a full layer is zero (see previous section), all the substrate layers behave as perfect (non deformed) layers from a diffraction viewpoint. On the contrary, as the mean deformation in the islands $\langle\varepsilon\rangle$ does not vanish before coalescence (see section I2) the islands behave as a mean deformed layer of in-plane lattice spacing :*

$$a = b_o(1+\langle\varepsilon\rangle). \qquad (12)$$

### I32/ Diffraction of the whole system: an interference average model

A first approximation of a dynamical scattering model valid for RHEED can be achieved by including in the classical kinematical theory the refractive effect of the average crystal potential [18,19], leading to the structure factor $F = \sum_j f_j \exp(-2i\pi\vec{s}.\vec{r})\exp(-i\Delta\Phi_j)$ where $s(s_x, s_y, s_z)$ is the scattering vector, $f_j$ the atomic scattering factor of the $j^{th}$ scatterer, and $\Delta\Phi_j=\Phi_j(V_n)-\Phi_j(0)$ with $\Phi_j(V_n) = 2\pi d_j \left( \sqrt{s_\perp^2 + \frac{2me}{h^2}V_n} + \sqrt{s_\perp'^2 + \frac{2me}{h^2}V_n} \right)$ a supplementary phase shift due to refraction [17]. In the previous equation $s_\perp$ and $s'_\perp$ are the surface regular components of the incident and diffracted vector, $d_j$ the distance from the j$^{th}$ scatterer to the surface and $V_n$ the crystal potential within the layer n to which belongs the j$^{th}$ scatterer. Following Horio et al [19] we assume a constant average crystal potential $V_n=V_0$ within each whole layer (n=2,3,..) but a coverage dependent potential for the growing layer (n=1): $V_{n=1} = \theta V_0$. For a pile-up of homogeneous layers the structure factor thus reads :

$$F = \sum_n F_n \exp(-i\Phi(V_n)) \qquad (13)$$

where $F_n$ is now the structure factor of the n$^{th}$ layer and the summation is performed for all the layers. In equation (13) $\Phi_j(V_n) = \Phi(V_n)$ is constant for all the scatterers that belong to the same layer n, with in fact $d_i$=nd where d is the inter-layers distance[4]. This last equation only means that the electron beam diffracted by the top layer (characterised by the inner-potential $V_1=\theta V_0$) interferes with the electron beams diffracted by the underlying layers characterised by the constant inner-potential $V_0$ (see figure 4).

It should be noted that, for a 1D crystal, the structure factor of a layer n (non deformed lattice spacing $a_o$) with lateral number of atoms $\Lambda$ and mean deformation $\langle\varepsilon\rangle$ so that $a = a_o(1+\langle\varepsilon\rangle)$ reads :

$F(\Lambda,\langle\varepsilon\rangle) = \sum_k f \exp(-2i\pi s_x ka) = f \frac{1-\exp(-2i\pi S_x \Lambda a_o(1+\langle\varepsilon\rangle))}{1-\exp(-2i\pi S_x a_o(1+\langle\varepsilon\rangle))}$. Thus for a pile-up of n non deformed substrate layers (lateral size $\Lambda$, mean deformation $\langle\varepsilon\rangle_\Lambda=0$) supporting a deformed layer (lateral size $\Lambda$, mean deformation $\langle\varepsilon\rangle_\Lambda=\langle\varepsilon\rangle$), (see section I31) the total structure factor (13) can be written as :

$$F_{tot} = F(\Lambda,\langle\varepsilon\rangle)(1+\tau^\theta \exp(-i\Phi(\theta))) + \tau^\theta \exp(-i\Phi(\theta))\sum_n F(\Lambda,0)\tau^n \exp(-in\Phi(\theta=1)) \qquad (14)$$

with $\quad \Phi(\theta) = 4\pi d\sqrt{s_\perp^2 + \frac{2me}{h^2}V_0\theta}$

---

[3] Furthermore we will see in section I33 that such satellites should be hidden by the supplementary satellites due to the mean correlation between the deposited islands that exist even in absence of misfit.

[4] Let us note since that for such a pile-up $s_z d_i = \Phi(V=0)$, the terms $\Phi(0)$ disappeared in (13)

where we choose $s_\perp = s'_\perp$ for the sake of simplicity and without any loss of generality. Let us note that we introduce some damping parameter $0<\tau<1$ for the complete layer and $\tau^\theta$ for the incomplete layer with coverage $0<\theta<1$. In the following, we note the usual phase shift $\Phi(\theta=1)$ as $\Phi$ since it is not coverage dependent.

Equation (14) then reads, with $s_x a_o = q$ the diffraction order (referred to the undeformed crystal):

$$F_{tot} = f \frac{\sin(\pi q \Lambda(1+\langle\varepsilon\rangle))}{\sin(\pi q(1+\langle\varepsilon\rangle))} (1+\tau^\theta \exp(-i\Phi(\theta))) + f\tau^{(1-\theta)} \exp(-i(\Phi(\theta)+\Phi)) \frac{\sin(\pi q \Lambda)}{\sin(\pi q)} \frac{1-\tau^n \exp(-in\Phi)}{1-\tau\exp(-i\Phi)} \quad (15)$$

In some peculiar cases, for instance for an odd number of layers and in out-of phase conditions ($\Phi = (2k+1)\pi$) and $\tau=1$ (no damping) $F_{tot}$ simply reads:

$$F_{tot} = f \frac{\sin(\pi q \Lambda(1+\langle\varepsilon\rangle))}{\sin(\pi q(1+\langle\varepsilon\rangle))} (1+\exp(-i\Phi(\theta))) \quad (15')$$

In equation (15'), the physical origin of the RHEED intensity oscillations is contained in the coverage dependent phase shift $\Phi(\theta)$ as depicted by Horio [19]. Moreover, the oscillations of the diffracted peak position originates in the coverage dependent mean deformation $\langle\varepsilon\rangle$. Furthermore this approximated expression (15') allows us to obtain an analytical expression for the position shift of the diffracted intensity maximum measured at half coverage $\theta=0.5$. For this purpose we derive the diffracted intensity $I = |F_{tot}|^2$ with respect to the diffraction order q. This derivative becomes equal to zero for the value $q = q_{max} = k(1+\langle\varepsilon\rangle_{0.5})^{-1} \approx k(1-\langle\varepsilon\rangle_{0.5})$ with k an integer and $\langle\varepsilon\rangle_{0.5}$ the mean deformation calculated for half coverage $\theta=l/L=1/2$. Using thus the analytical expression of the mean deformation given in table II, the position shift of the diffracted peak calculated at half coverage reads (with k=1 since we only consider the first order peak):

$$\Delta q = \frac{8}{\pi} \frac{a}{L} Km \ln(\sin(\pi a/L)) \quad (15'')$$

It is thus easy to see that the shift of the peak position depends linearly on the misfit and varies with the nucleation density 1/L. It should be noted that equation (15'') only gives an overestimation of the true shift. Indeed when taking into account some electron absorption ($\tau \neq 1$), $F_{tot}$ given by (15) contains in fact two contributions: an island contribution depending on the mean deformation in the island $\langle \varepsilon \rangle$ and a substrate contribution centred at integer values of q as suggested by [8]. Generally, because of the poor resolution of the RHEED detector, the two contributions are not resolved and only one diffracted peak is recorded whose position is intermediate between the one given by (15'') and the regular one due to the non deformed substrate. In other words in order to obtain the true position-shift the overestimated position-shift given by (15'') should be multiplied by an unknown coefficient depending on the electron absorption coefficient τ.

We can use the total structure factor given by (15) to plot the diffracted intensity $I(q) = F_{tot} F_{tot}^*$ as a function of the diffraction order q for various values of coverage θ. In figure 5 we plot I(q) for various values of the coverage θ and for L=20 where the mean deformation in the islands is large (see fig. 3b). (We also take Λ=20 giving a broadening of the diffracted peak compatible with experiment). As RHEED intensity oscillations are phase dependent with the incident angle [19], the incidence angle is adjusted in the calculation reported in figure 5 in order to obtain a minimum of intensity for roughly zero coverage (furthermore we take $V_o$=10 V and τ=0.5 as it should be for electrons diffraction [17,19]). A single diffraction peak is obtained, whose intensity and position vary with the coverage θ as reported in the various experiments [3-7]. In figure 6 is reported the shift of the maximum of intensity (Δq) as a function of the coverage calculated for L=20 and L=100 and m=5%. The position oscillations clearly pass through a maximum at roughly θ=0.4-0.5. As we will see in the following the smaller is L, the greater the displacement. Furthermore let us note that, as in experiments (see [3-7] and also our experimental results on figure 9), the oscillation is not symmetric with respect to the maximum value θ≈0.4. The intensity value at zero coverage only depends on the relative fraction of substrate felt by the electrons or in other words on the factor $\tau^n$ (for fixed diffusion factor f and lateral size L). In figure 7 we report the position shift of the diffraction peak as a function of the misfit for L=20 (fig 7a) and of the nucleation density 1/L for m= 5% (fig 7b). We also report on the same figure the overestimated shift obtained from the approximated expression (15''). *We see that, as predicted by (15''), the amplitude of the oscillation of position depends linearly on the misfit m (more precisely on the product Km). The dependence of the oscillation of position with the nucleation density is more complicated. Nevertheless in a large domain of nucleation density it can be considered* *to depend roughly linearly on the nucleation.* Thus one can conclude that in a large range of nucleation density:

$$\Delta q \propto Km/L \qquad (16)$$

### I33/ Comments about FWHM oscillations

The simple analytical approach we develop allows to describe intensity and position oscillations as well, but not the oscillations of full-width at half maximum (FWHM) reported in [4-7] and usually attributed, but without any proof, to a size effect (Debye effect). In fact, this FWHM oscillations may be explained by taking into account the diffuse scattering which is mixed with the specular diffraction in RHEED experiment. Indeed, if the contribution of the diffuse scattering in X-Ray or neutron diffraction [26] is quite easy to separate from the specular scattering (because of the good resolution in the reciprocal space that can be achieved with these techniques), this is however not so easy with regular electron diffraction apparatus. The diffuse and specular scattering are thus often mixed into one peak. The determination of both contributions may be achieved by measuring only one peak or spot as shown a long time ago by Henzler with the LEED technique [23], and more recently by Stroscio et al. [27,28] and Dulot et al [20,21] using RHEED in the case of the Fe homo-epitaxy. More precisely the intensity of these satellites varies with the coverage (in opposite phase with respect to the Bragg peak) whereas the FWHM of both satellites and Bragg peak remains roughly constant whatever coverage. In absence of a good enough resolution (due to the apparatus) the diffracted profile can be fitted by a single Lorentzian peak whose FWHM then oscillates with coverage as numerically shown in [21]. *In other words, the FWHM oscillations are an artefact due to the poor resolution of the RHEED instrument*.

Such satellites have been reported and extensively studied by many authors in the case of diffraction by a surface with random terraces distribution [22-24]. Analytical calculations have been performed for one dimensional geometry and in the framework of a two-level model [22,24]. In appendix III, following [22-24] and in the case of a Lorentzian distribution[5] of terrace lengths with same variance in both level, we show that the first satellites positions are $s_x = \pm 1/L$ for $\theta = 1/2$ with a weak variation of position when the coverage varies in the range $0.1 < \theta < 0.9$. The satellites intensities roughly vary as $s_x^{-2}$ leading to a strong attenuation of the satellites intensity with the diffraction order. It can be numerically checked

---

[5] In fact, the position of the first satellite does not depend upon the form of the distribution probability provided the variance σ does not exceed 20% of the mean length [24]

that, as in experiment, the Bragg peak and the first satellites intensities oscillate with coverage but in opposite phase and with constant FWHM. *It is important to note that at $\theta = 1/2$ where in this simple two-level model the Bragg peak intensity vanishes (see appendix III in the out-of-phase condition of the RHEED experiment), the distance in between the satellites, analogous to the FWHM of a mean poorly-resolved peak, is 2/L and thus gives access to the nucleation density 1/L as :*

$$FWHM(\theta = 0.5) \propto 1/L \qquad (17)$$

## I4/ Conclusion

The main conclusions of the theoretical part are the following :

**(1)** The in-plane lattice spacing oscillations originate from the edge elastic relaxation of 2D islands even in the case of homoepitaxy where the active misfit only originates from size effect.

**(2)** Owing to the elastic relaxation the displacement of the island edges is maximum at half coverage even when the growth takes place at constant step density.

**(3)** The mean deformation in the island reads $\langle \varepsilon \rangle = Kmf(l, L)$ where K is the relative rigidity, m the so-called active misfit and $f(l,L)$ a function of the island size $l$ and the nucleation density 1/L.

**(4)** Because of the interference between the electrons scattered by the growing layer and the underlying layers, the RHEED rod-spacing oscillations does not give directly access to the in-plane lattice spacing oscillation.

**(5)** The displacement of the RHEED rods measured at half coverage varies linearly with the product Km and varies roughly linearly with the nucleation density.

**(6)** The FWHM measured at half coverage gives access to the nucleation density 1/L, if the experiments are *strictly* performed in anti-bragg geometry.

Consequently, the detected relaxation is only a small part of the true relaxation. In other words, as we have shown in another paper [8] (but from a phenomenological view-point), there is some instrumental distorsion of the rod displacement with respect to the true deformation. In [8] we proposed a phenomenological expression of the transfer function by which the experimental relaxation has to be multiplied to obtain the true relaxation. In order to establish such a transfer function we assumed that the RHEED intensity could be written as the sum of two terms. A first term originates from the non deformed substrate and a second one from the deformed island but furthermore modulated in intensity by some oscillatory function of coverage to account for the usual RHEED intensity oscillations. All these seminal ideas are now justified, but when the interferences in between the growing layer and its underlying substrate are properly considered there is no more simple transfer function. In other words the experimental results of rod displacements have to be fit by (15) to extract the true mean deformation. Nevertheless let us note that stricto-sensu (15) is only valid for 1D islands. Finally, deducing the true relaxation effect from the measurements is also complicated by their dependence on the nucleation density.

## II/ Comparison with experimental data:

Since the distance in between islands (L) is an open parameter in the theoretical relations of section I, a comparison between experiments and theory may be convincing only if the nucleation density is known. Consequently, after a short description of the experiments (section II1), we show that, in agreement with section I33, the measured FWHM gives access to the nucleation density in good approximation (section II2). By varying the nucleation density using impurities adsorption or by varying the substrate temperature, the theoretical prediction are thus tested in section II3.

### II1/ Description of experiments

Let us stress on four points:

**(i)** The epitaxial films are grown by MBE on MgO substrates for typical growth rates ranging from 1 to 10 Å/min and substrate temperatures from 300 to 1100K. The in-plane lattice spacing variations are measured by using RHEED. The experimental details concerning the growth of the epitaxial films and the measurement of RHEED intensity, FWHM and in-plane lattice spacing variations ($<a_{//}>$) were already given in a previous paper [7]. We just want to give the general method and some additional information about the calibration of the $<a_{//}>$ and FWHM quantities. The measurement is performed by recording during the deposit a profile on a RHEED pattern perpendicular to the streaks and including the (0-1) (00) and (01) streaks. This profile is fitted by using three lorentzian peaks. We thus get the intensity, FWHM and position of each peak. The inverse of the distance between two peaks is converted into experimental in-plane lattice spacing (it is not the true lattice spacing (see previous section remark 4)), which is calibrated by fixing the initial distance between two streaks of the substrate surface to unity. In order to improve the sensitivity on the distance measurement, we always measure the distance between (0-1) and (01) peaks. Consequently, we only report the relative variations of the in-plane lattice spacing during the growth compared to this initial substrate in-plane lattice. *We thus not obtain an absolute measurement of the in-plane lattice spacing*, since the in-plane lattice spacing of the initial surface could be different from the bulk, especially when surface reconstruction takes place. However, the difference between this surface lattice spacing and the bulk, if it exists, is small. Consequently the FWHM is calibrated by considering that the distance between two streaks on RHEED patterns of the substrate is proportional to the inverse of the bulk distance.

**(ii)** As shown in section I33, the determination of the nucleation density by using the FWHM of RHEED rods is in theory possible, but with the condition that we are looking at a peak in anti-bragg condition. It should be noted that the anti-bragg geometry is not necessarily the same for a (00) peak as for a (01) peak. In other words, if we are looking at a profile that crosses the (00) streak in specular geometry with an incident angle corresponding to the anti-bragg condition for this (00) streak, it is not necessarily the same for the (01) peak. For instance, if the experiment is performed on a bcc or fcc (001) surface in the [10] azimuth of the square lattice, an anti-bragg condition on the (00) peak is a bragg condition on the (01) peak. On the contrary, the anti-bragg condition is valid for both (00) and (01) streaks in specular geometry in the [11] azimuth of the square lattice. This important point is taken into account in our experiments. In that case, the nucleation density is given by the square of $FWHM/4\pi$. However the FWHM of a RHEED streak results from the convolution of the diffracted peak and instrumental peak coming from the limited resolution of the apparatus. This resolution is limited by the angular divergence of the electron beam but also by the resolution of our CCD camera. The total FWHM is thus the addition (because a diffracted pattern is in reciprocal space) of the FWHM of the apparatus and the FWHM coming from the surface. The total FWHM due to the apparatus is easy to determine by using a surface with large terraces. A STM analysis shows that the size of the terraces on the (001) V and Fe buffer layers used in this study are larger than 500Å [25]. In that case, the FWHM of the peaks measured on the initial RHEED patterns is essentially due to our system resolution. Consequently, getting the FWHM coming from the surface is easy : the initial FWHM measured before the deposit is started is subtracted to the total FWHM curve versus time deposition.

**(iii)** For studying the nucleation density effect on <$a_{//}$>, the nucleation density has to be varied. In practice the nucleation density can be varied in three ways, (i) by changing the incoming flux, (ii) by varying the substrate temperature, (iii) by adsorbing some impurities on the surface before the growth, which act as nucleation centres. The variation of the substrate temperature is used in the case of the homoepitaxy of (001) V, Fe and Nb. In the case of heteroepitaxy, however, the variation of the nucleation density by varying the temperature is not a suitable method since interdiffusion often takes place. We consequently use the adsorption of oxygen to vary this nucleation density. The oxygen contamination of the surface is achieved by heating the buffer layer at elevated temperatures (applied for Mn/Fe) or by exposing the surface to $O_2$ (applied for V/V(001) ). This oxygen adsorption is controlled by Auger spectroscopy. However, we do not calibrate the oxygen surface concentration, since quantitative Auger analysis is always difficult and this is not necessary for our purpose.

**(iv)** Finally, the relative rigidity K we define is the one of the substrate with respect to the one of the 2D island. In the case of heteroepitaxy this relative rigidity cannot be the same if the substrate is B or if the substrate is a composite of n pseudomorphous layers of A deposited on B. The relative rigidity thus may vary from one deposited layer to the other. We believe that this effect is of second order but in the following we will only discuss the case of the first oscillation where K is then the real relative rigidity of the substrate B with respect to the island A.

II2/ FWHM versus nucleation density

According to the discussion in section I33, the maximum FWHM near half coverage is first assumed to be proportional to the inverse distance between 2D islands (keeping in mind that the experiments are performed in anti-bragg geometry). From an experimental view-point we measure systematically the variation of the FWHM of the (0-1) and (01) peaks (but not of the (00) peak) and experimentally show that the FWHMs measured on these peaks (substracting the width of the electron beam and detector) effectively give values close to the nucleation densities observed by STM. For this purpose we report in figure 8 the variation of the nucleation density calculated by using the maximum FWHM of the (01) peak with temperature for the homoepitaxial Fe, V and Nb systems. In the case of Fe, the results are compared to the data obtained by Stroscio et al on Fe(001) using STM [27,28]. The agreement between STM and FWHM nucleation density determination is very good, in agreement with results reported in [21]. In the case of V and Nb, it should be noted that the variation of the nucleation density strongly depends on the initial oxygen surface coverage, as shown in the following in the case of V. The results reported in figure 8 correspond to an initial surface reconstructed in 5x1 with a constant oxygen coverage. Finally, it should be noted that two slopes are observed in the Arhenius plots of V and Fe. This behaviour is due to the well known change of the critical size of stable nuclei [27,28]. This is another proof that *the* *maximum FWHM measured on the (01) peak in the corresponding anti-bragg geometry gives a good approximation of the nucleation density.*

**II3/ <$a_{//}$> oscillations amplitude versus nucleation density**

The estimation of the nucleation density now allows us to understand why the amplitude of the <$a_{//}$> oscillations may vary from one experiment to another for a given system as mentioned in [7]. We show for instance in figure 9 two experiments performed on the systems Mn/Fe(001), V/Fe(001) and Co/Cu(001). The difference in these experiments is the amount of oxygen detected on the surface before each experiment. We first discuss the case of Mn/Fe(001). For very small amount of oxygen (undetectable by AES), the amplitude of the in-plane lattice oscillations is around 0.8%. However, it drops up to more than 3% when oxygen is present on the surface. Simultaneously, we observe that the FWHM increases. The cases of V/Fe(001) and Co/Cu(001) systems are also very interesting. Indeed, for small nucleation densities (large islands at half coverage), no relaxation effects are detected in both cases. However, some detectable relaxation effects are recorded when the FWHM (proportional to the nucleation density) increased. Finally, a similar behaviour is observed in the case of the V(001) epitaxy [25].

These results are simply explained by the fact that oxygen atoms play the role of nucleation centres and multiply the number of stable nucleus. Consequently, the nucleation density is increased, and the 2D islands size at half coverage is decreased. On a quantitative view-point, we observe for five systems a linear variation of the amplitude of the first in-plane lattice spacing oscillation with the amplitude of the first FWHM oscillation (fig.10). This means that the *<$a_{//}$> oscillations amplitude is proportional to the inverse distance between 2D islands* as numerically shown in figure 7b. Let us underline that such linear dependence over all the nucleation density range is not in perfect agreement with theoretical expressions of section I32. We believe that such a discrepancy comes from the reduced dimensionality of the model (1D) compared to the experimental data (2D growth).

II4/ Elastic relaxation and effective misfit

In our previous paper [7], a quantitative comparison of the relaxation amplitude with the misfit for all the studied systems was not possible. Indeed, the detected relaxation effect depends on the size of the 2D islands, and consequently on the nucleation density, which was not a parameter under control. Since now the nucleation density is available, we propose to reconsider this particular point. Since we have shown that the maximum shift of the rod can be roughly written $\Delta q_{max} \propto Km/L$ (see equation (16)) one can obtain by simple differentiation of (16):

$$Km \propto \Delta a_{max}/FWHM \qquad (18)$$

where $\Delta a_{max}$ is the maximal detected relaxation and FWHM given by (17) is the full width at half coverage. For heteroepitaxial system with misfit greater than the percent, the active misfit m must roughly be equal to the natural misfit $m_o$. In order to verify relation (18) we thus have to plot $Km_o$ as a function of $\Delta a_{max}/FWHM$ (proportional to the slope of the straight lines in figure 10) for the various systems under study. Nevertheless it is not trivial since for these systems the structure of the material A deposited on B is generally not its regular structure stable in regular conditions of pressure and temperatures (metastable structure). For instance, Ni is known to grow on Fe (2.866Å) in its bcc structure (a=2.773Å), Co on Cu (3.615Å) and Ni (3.52Å) in its fcc structure (3.545Å), Fe on Cu in its fcc structure (3.59Å), Mn on Fe in its bcc structure (2.92±0.03Å). Thus, since to the best of our knowledge the elastic constants of these structures are not available, the relative rigidity K appearing in (18) cannot be estimated ! The case of V homoepitaxy is peculiar since in this case K=1 and $m_o$ should be zero whereas in figure (10) the slope of the corresponding straight line, proportional to Km, is far from being zero. Such behaviour could be attributed to the difference between the natural misfit $m_o$ and the active misfit m we define in section I1. Nevertheless, we have shown in a previous paper [25] that the reconstructed vanadium surface has a lattice spacing 6% larger than pure V. This corresponds to a misfit of 6%, which explains why so large surface relaxation effects are observed in this special case of homoepitaxy. In summary, we can say that, in order to thoroughly check the theoretical expressions, one has to study some other systems for which the anisotropic elastic constants and natural misfit are well known. In the case of homoepitaxy, the true natural misfit, that means taking into account the eventual surface reconstruction, has also to be perfectly known.

## III/ Conclusion and perspectives

In this paper, we unambiguously demonstrate the elastic origin of the in-place lattice spacing oscillations. We find some analytical equations driving the oscillation amplitude that can be experimentally checked. Though based on simple arguments in 1D geometry, the calculations are in good semi-quantitative agreement with experimental data obtained on many metallic systems (though 2D). Another interesting result is the connection between the detected relaxation effect and the nucleation density, both on the theoretical and experimental points of view. As a consequence, two in-plane lattice spacing experiments cannot be compared without knowing the nucleation density. This certainly explains some apparent discrepancies in the literature. For example, the fact that we do not detect any relaxation effect in the case of Cu(001) epitaxy [7], on the contray to Fassbender et al [5], may probably be explained by some differences in the nucleation densities in both experiments.

Obviously some other experiments have to be performed in order to check the validity of the assumptions used in the calculations. First, it seems important to check other systems and in particular homoepitaxial systems for which a foreign adsorption may play a role on the nucleation density and the active misfit as well (by incorporation or simple adsorption). Particularly, the lateral size effect contained in the $c_1/l$ contribution to the effective misfit (see relation (5)) should play an important role especially when the 2D islands are laterally small at the beginning of the nucleation process (where nevertheless the nucleation density is not constant). Second, it should be interesting to use the analytical equations obtained in section I in order to extract (by a fitting procedure) the true mean deformation from experimental data on rod shift and compare it to relaxation measurements that may be obtained by other techniques, like STM for instance. Likely one must go beyond this 1D theory and then develop a more complete approach. Third, the effect of some experimental parameters that appear in the analytical formulation have to be checked. It is for example the case of the incident angle influence.

**Acknowledgements:**
Professor R.Kern is acknowledged with gratitude for very fruitful discussions.

## Appendix I:

For the sake of simplicity we only consider a simple cubic crystal described by a pair potential $\Phi(r)$. The cohesion energy per atom $U^{3D}$ of a 3D crystal (parameter $a^{3D}$) thus reads $U^{3D} = \frac{1}{2}\left[6\Phi(a^{3D}) + 12\Phi(a^{3D}\sqrt{2})\right]$ where we only consider 6 first and 12 second neighbours. At the same the cohesion energy per atom $U^{2D}$ of a 2D (100) free standing crystal (parameter $a^{2D}$) reads with the same approximation $U^{2D} = \frac{1}{2}\left[4\Phi(a^{2D}) + 4\Phi(a^{2D}\sqrt{2})\right]$. At equilibrium there is furthermore: $\partial U^i / \partial a^i = 0$ where i=3D or 2D so that $\Phi'(a_0^{3D}) = -2\sqrt{2}\Phi'(a_0^{3D}\sqrt{2})$ and $\Phi'(a_0^{2D}) = -\sqrt{2}\Phi'(a_0^{2D}\sqrt{2})$ respectively when $a_0^i$ are the equilibrium parameters. ($\Phi'(r)$ and $\Phi''(r)$ are the first and second derivative of the potential respectively).

*For an infinitely large 2D crystal*, one can define $a_0^{2D} = a_0^{3D}(1+\varepsilon_{xx}^\infty)$ where $\varepsilon_{xx}^\infty$ is thus the relative deformation of the 2D infinite crystal with respect to the 3D one. We can thus write the last equation under the form $\Phi'(a_0^{3D}(1+\varepsilon_{xx}^\infty)) = -\sqrt{2}\Phi'(a_0^{3D}(1+\varepsilon_{xx}^\infty)\sqrt{2})$ that gives after development (to first order in strain):

$$\varepsilon_{xx}^\infty = \frac{1}{2} \frac{\Phi'(a_0^{3D})}{a_0^{3D}\left[\Phi''(a_0^{3D}) + 2\Phi''(a_0^{3D}\sqrt{2})\right]} \qquad \textbf{(i)}$$

\* *For a 2D ribbon of finite lateral size* $l_x$ ($l_y$ *being infinite*[6]) there is to consider the step energy ρ of the edges (parallel to axis y): $\rho = \frac{1}{2a_y}\left[\Phi(a_x) + 2\Phi\left(\sqrt{a_x^2 + a_y^2}\right)\right]$ where we make a distinction between $a_x$ and $a_y$ the crystallographic parameter perpendicular and parallel to the edges respectively, because of the stressed edges $a_x \neq a_y = a_0^{2D}$. Usually surface stress and surface energy are connected by way of the Shuttleworth equation [29]. Here we generalise this former equation by defining a step stress r connected to the step energy ρ by some Shuttleworth relation $r = \rho + \partial\rho/\partial\varepsilon_{yy} = \rho + a_y\,\partial\rho/\partial a_y$ so that $r = -\frac{1}{2}\frac{\Phi'(a_0^{2D})}{a_0^{2D}}$. Defining then a 2D Young modulus as $2Y^{2D} = d^2U^{2D}/d(a_0^{2D})^2 = \frac{1}{a_0^{2D}}\left[\Phi''(a_0^{2D}) + 2\Phi''(a_0^{2D}\sqrt{2})\right]$ one can define similarly to (2) the deformation due to the unique size effect as $\varepsilon_{xx}^{l_x} = \frac{1}{Y^{2D}}\frac{r}{l_x}$ (there is no more Poisson effect for 2D crystal) so that

$$\varepsilon_{xx}^{l_x} = -\frac{1}{2}\frac{\Phi'(a_0^{2D})}{\left[\Phi''(a_0^{2D}) + 2\Phi''(a_0^{2D}\sqrt{2})\right]}\frac{1}{l_x} \qquad \textbf{(b)}$$

Summing up the two effects (a) and (b) the strain of a 2D ribbon of lateral size l with respect to a 3D crystal reads:

$\varepsilon(a,l) = \varepsilon_{xx}^{\infty} + \varepsilon_{xx}^{l_x}$ which is of the form $\varepsilon(a,l) = C_o + C_1/l$

---

[6] Our model in the following is a 1D model of islands

## Appendix II

Let us recall the classical formulation of diffraction by an imperfect crystal [30,31]. For a perfect crystal where each unit cell repeats periodically, all unit cells have the same diffraction properties described by the structure factor $F$. For an imperfect crystal, derived from the same lattice, each unit cell has its own structure factor $F_n$. The diffracted intensity of the imperfect crystal thus reads:

$$I = \sum_n \sum_{n'} F_n F_{n'}^* \exp[-2i\pi \vec{s} \cdot (\vec{x}_n - \vec{x}_{n'})] \qquad \textbf{(i)}$$

Where $x_n$ characterises the position of the $n^{th}$ cell

In the perturbation theory of imperfect lattice [30,31], one introduces an average structure factor over N cells $\langle F \rangle = \frac{1}{N} \sum_n F_n$ and its perturbation under the form $F_n = \langle F \rangle + \varphi_n$

The diffracted intensity thus reads [30]

$$I = \langle F \rangle \langle F^* \rangle \sum_p V \exp(2i\pi \vec{s} \cdot \vec{x}_p) + \sum_p V \langle \varphi_n \varphi_{n+p} \rangle \exp(2i\pi \vec{s} \cdot \vec{x}_p) \qquad \textbf{(ii)}$$

where the first term describes the intensity scattered by the average lattice (with s a reciprocal vector) and the second term which depends on the correlation function $\langle \varphi_n \varphi_{n+p} \rangle$ describes the deviation from the average lattice.

In our case, we only consider one atom per unit cell of structure factor F=f, and we calculate the average structure factor as :

$$\langle F_n \rangle = \langle f \exp(-2i\pi \vec{s}.\vec{u}_n) \rangle \approx f \langle 1 - 2i\pi \vec{s}.\vec{u}_n - 2\pi^2 (\vec{s}.\vec{u}_n)^2 \rangle \qquad \text{(iii)}$$

where, f is the diffusion factor, $u_n$ the displacement field of the $n^{th}$ atom (see (9)) and where we have neglected higher powers of ($\vec{s}.\vec{u}_n$). Since the mean displacement $\langle \vec{u}_n \rangle$ is zero, equation (iii) thus becomes

$$\langle F_n \rangle = f \left[ 1 - 2\pi^2 s^2 \langle \vec{u}_n^2 \rangle \right] \qquad \text{(iv)}$$

and

$$\varphi_n = F_n - \langle F_n \rangle = f \left[ \exp(-2i\pi \vec{s}.\vec{u}_n) - 1 + 2\pi^2 s^2 \langle \vec{u}_n^2 \rangle \right]$$

We use then a Fourier development of the displacement field (9)

$$\vec{u}_x^B(x) = -\frac{2m}{\pi} \sum_{N=1} \frac{\sin(\pi N \theta)}{N} \sin(2\pi N \vec{k}.\vec{x}_n) \vec{i} \qquad \text{(v)}$$

where $\vec{i}$ is a unit vector perpendicular to the ribbons (see figure 1), k=1/L the wave vector of the displacement wave and $\theta = l/L$ the coverage.

One can thus calculate $\langle \vec{u}_n^2 \rangle = \left( \frac{2m}{\pi} \right)^2 \left\langle \sum_{N=1} \left( \frac{\sin(\pi N \theta)}{N} \right)^2 (\sin(2\pi N \vec{k}.\vec{x}_n))^2 \right\rangle$ or owing to sinus average properties

$$\langle \vec{u}_n^2 \rangle = \left( \frac{2m}{\pi} \right)^2 \frac{1}{2} \sum_{N=1} \left( \frac{\sin(\pi N \theta)}{N} \right)^2 \qquad \text{(vi)}$$

We thus obtain the correlation function $\varphi_n \varphi_{n+p}^* = f^2 [-2i\pi \vec{s}.\vec{u}_n][2i\pi \vec{s}.\vec{u}_{n+p}]$ whose average value is

$$\langle \varphi_n \varphi_{n+p}^* \rangle = f^2 4\pi^2 s^2 \langle \vec{u}_n.\vec{u}_{n+p} \rangle. \qquad \text{(vii)}$$

Then writing $\vec{x}_{n+p} = \vec{x}_n + \vec{x}_p$ one obtains

$$\langle \vec{u}_n.\vec{u}_{n+p} \rangle = \left( \frac{2m}{\pi} \right)^2 \left\langle \sum_N \left( \frac{\sin(\pi N \theta)}{N} \right)^2 \sin(2\pi N \vec{k}.\vec{x}_n) \sin(2\pi N \vec{k}.(\vec{x}_n + \vec{x}_p)) \right\rangle$$

which gives after development of $\sin(2\pi N \vec{k}.(\vec{x}_n + \vec{x}_{n+p}))$ with $\vec{x}_n = n\vec{a}$,

$$\langle \varphi_n \varphi_{n+p}^* \rangle = f^2 4\pi^2 q^2 \left( \frac{2m}{\pi} \right)^2 \frac{1}{2} \sum_N \left( \frac{\sin(\pi N \theta)}{N} \right)^2 \cos(2\pi N \vec{k}.\vec{x}_p) \qquad \text{(viii)}$$

where we write $\vec{s}.\vec{a} = q$ the diffraction order. The diffracted intensity (i) thus reads:

$$I = f^2 \sum_p \exp(2i\pi \vec{s}.\vec{x}_p) + \sum_N \sum_p f^2 4\pi^2 q^2 \left( \frac{2m}{\pi} \right)^2 \frac{1}{2} \left( \frac{\sin(\pi N \theta)}{N} \right)^2 \cos(2\pi N \vec{k}.\vec{x}_p) \exp(2i\pi \vec{s}.\vec{x}_p)$$

or when using Euler formulae:

$$I = f^2 \sum_p \exp(2i\pi \vec{s}.\vec{x}_p) + \sum_N \sum_p f^2 4\pi^2 q^2 \left(\frac{2m}{\pi}\right)^2 \frac{1}{2}\left(\frac{\sin(\pi N\theta)}{N}\right)^2 \left[\exp(2i\pi(\vec{s}+N\vec{k}).\vec{x}_p) + \exp(2i\pi(\vec{s}-N\vec{k}).\vec{x}_p)\right]$$

# Appendix III

For one dimensional geometry and in the framework of a two-level model the intensity diffracted by a surface with random terrace distribution reads in out of phase conditions ($2\pi s_z d = \pi$) [22,24]: $I = I_{bragg} + I_{diffus} \propto (1-2\theta)^2 \delta(s_x) + C(s_x)(1-\delta(s_x))$. This intensity is a delta function term (Bragg peak) to which is added a non singular term depending on the reduced correlation function of the distribution of terraces $C(s_x)$. This last term can also be written: $I_{diffus} \propto \frac{\theta}{s_x^2} Re\left[\frac{(1-P_1(s_x))(1-P_2(s_x))}{1-P_1(s_x)P_2(s_x)}\right]$ where $P_i(s_x)$ is the Fourier transform of the probability $P_i(x)$ of finding a terrace of length x at the $i^{th}$ level and where Re(Y) is the real part of the complex function Y. Obviously these expressions have to be convoluted by an instrumental function.

If one chooses a Lorentzian distribution of terraces with the same variance in the two levels that means the probabilities $P_i(x) = ((x-l_i)^2 + \sigma^2)^{-1}$ having the Fourier transform $P_i(s_x) = \rho_i \exp(-is_x l_i)$ with $\rho_1 = \rho_2 = \frac{\pi}{\sigma}\exp(-s_x \sigma) \equiv \rho$, the previous equation becomes with $l_1 = L + \Delta L$ and $l_2 = L - \Delta L$

$$I_{diffus} \propto \frac{\theta}{s_x^2}\left[\frac{1-\cos(s_x \Delta L)}{1-\rho^2 - 2\rho\cos(s_x L)} + \frac{1+\cos(s_x \Delta L)}{1-\rho^2 + 2\rho\cos(s_x L)}\right]\left(\frac{1-\rho^2}{2}\right)$$ which is analogous to the

equation (36) of [24] but corrected from a typing error.

It is thus easy to develop this last equation around $\theta = 0.5$ that means around $\Delta L = L(1-2\theta) = 0$ up to the second order so that

$$I_{diffus} \propto \frac{\theta}{s_x^2}\left[\frac{1-\rho^2}{1-\rho^2+2\rho\cos(s_x L)}\right] + O_2(\Delta L)$$ having a maximum for $s_x = \pm(2k+1)/L$. We verify by numerical calculations that the maximum of the first satellite $s_x = \pm 1/L$ remains roughly the same for $0.1 < \theta < 0.9$. This is essentially due to the absence of a first order in the just-above development.

# Figure captions

**Figure 1:** Sketch of the model : set of periodic (L apart) infinite ribbons (width l) of material A deposited onto a mismatched substrate B.

**Figure 2:** Normalised displacement fields $u_x^i(x)/(Kma/\pi)$ where $u_x^0(x)$, $u_x^{Int}(x)$ and $u_x^B(x) = u_x^0(x) + u_x^{Int}(x)$, are the displacement field in absence of elastic relaxation, the elastic relaxation contribution and the total field respectively (see equation (9) and table I). Figures 2a, 2b and 2c correspond to a coverage θ=0.1, 0.5 and 0.99 respectively and are calculated for a coalescence size L=100. Owing to the elastic relaxation, the greater the coverage is, the closer the border of the islands and the weaker the displacement field.

**Figure 3: a)** Normalised displacement $u_x^i/(Kma/\pi)$ calculated at the island edge $x = l/2$ as a function of coverage θ for L=100. $u_x^0$, $u_x^{Int}$ and $u_x^B = u_x^0 + u_x^{Int}$, obtained from equation (9) and table I, are the displacement field in absence of elastic relaxation, the elastic relaxation contribution and the total field respectively. Notice that $u_x^B$ reaches a maximum at half coverage. **b)** Normalised mean deformation $\langle\varepsilon\rangle/(Kma/\pi)$ as a function of coverage θ for various values of the coalescence size L.

**Figure 4:** Model used to calculate the diffracted intensities: the first growing layer is considered as a full weakly deformed layer whose inner potential varies with coverage, the underlying layers being considered as full non deformed layers (see the end of section I31). The electron beams diffracted by each layer interfere each other.

**Figure 5:** Diffracted intensity calculated from (15) as a function of q for various values of coverage θ=0.1, 0.2, 0.4, 0.5, 0.6, 0.8 (with α=1.4°, n=5; τ=0.5 and V=10 V). Notice the intensity oscillation and the position oscillation as well.

**Figure 6:** Shift of the diffracted intensity maximum versus coverage calculated for m=5%.: Triangles: L=100 , squares: L=20.

**Figure 7: a**) Shift of the diffracted intensity maximum versus misfit calculated for L=20**. b)** Shift of the diffracted intensity maximum versus nucleation density calculated for m=5%. In a large domain of nucleation density the variation is linear. (The diffracted intensity maximum are calculated for θ=0.5). In both cases the fat line corresponds to the overestimated shift calculated by (15'') whereas the squares linked by the thin line corresponds to the true shift calculated from (15) (see text).

**Figure 8:** Variation of the nucleation density (calculated from the RHEED peak FWHM) with the substrate temperature, and comparison to the nucleation density obtained by Stroscio et al. [27,28] using STM.

**Figure 9:** FWHM and in-plane lattice spacing oscillations (IPLOSs) obtained without (bottom) and with initial oxygen surface concentration (top) during the growth of Mn on (001) Fe (left), V on (001) Fe (middle), and Co on (001) Cu (right). The FWHM is increased in presence of oxygen due to the increase of nucleation centres. As a consequence, the 2D islands size decreases and the relaxation effect occurring at the islands edge is easier to detect.

**Figure 10:** Maximum variation of the detected in-plane lattice spacing oscillations (IPLSOSs) versus FWHM for Mn, Ni and V on Fe(001), Co on Cu(001) and V on V(001).

Tables

| $u_x^0(x)$ | $u_x^{Int}(x)$ | $u_x^B(x)$ |
|---|---|---|
| $\dfrac{a}{2} Km \ln\left[\dfrac{x-l/2}{x+l/2}\right]$ | $\dfrac{a}{2} Km \ln\left[\dfrac{\sin(\pi(x-l/2)/L)}{\pi(x-l/2)/L} \Big/ \dfrac{\sin(\pi(x+l/2)/L)}{\pi(x+l/2)/L}\right]$ | $\dfrac{2a}{\pi} Km \ln\left[\dfrac{\sin((\pi/L)(x-l/2))}{\sin((\pi/L)(x+l/2))}\right]$ |

***Table I:*** *Analytical expressions of the displacements fields at the surface z=0. The term $u_x^0(x)$ is the elastic field in absence of any elastic interaction and the term $u_x^B(x) = u_x^0(x) + u_x^{Int}(x)$ the total elastic field where $u_x^{Int}(x)$ is the contribution of the elastic interactions. Notice that these expressions are valid for $-\infty < x < \infty$ excepted at $x = \pm l/2$ where some local divergences occur because of the use of point forces.*

| $\langle\varepsilon\rangle^0 = \dfrac{4a}{\pi l} Km \ln \dfrac{a}{l}$ | $\langle\varepsilon\rangle^{Int} = \dfrac{4a}{\pi l} Km \ln\left[\dfrac{l}{a}\dfrac{\sin(\pi a/L)}{\sin(\pi l/L)}\right]$ | $\langle\varepsilon\rangle = \dfrac{4a}{\pi l} Km \ln\left[\dfrac{\sin(\pi a/L)}{\sin(\pi l/L)}\right]$ |
|---|---|---|
| $\langle\varepsilon\rangle^0_{out} = \dfrac{4a}{\pi l} Km \dfrac{l}{L-l} \ln\left[\dfrac{\sin(\pi l/L)}{\sin(\pi a/L)} \dfrac{L+l}{L-l} \dfrac{a}{l+a}\right]$ | $\langle\varepsilon\rangle^{Int}_{out} = \dfrac{4a}{\pi l} \dfrac{l}{L-l} Km \ln\left[\dfrac{L-l}{L+l} \dfrac{l+a}{a}\right]$ | $\langle\varepsilon\rangle_{out} = \dfrac{4a}{\pi l} Km \dfrac{l}{L-l} \ln\left[\dfrac{\sin(\pi l/L)}{\sin(\pi a/L)}\right]$ |

***Table II:*** *Analytical expressions of the mean deformations inside and outside the islands, in absence of any elastic interaction (subscript 0), only due to interactions (subscipt Int) and their sum.*

Figure 1

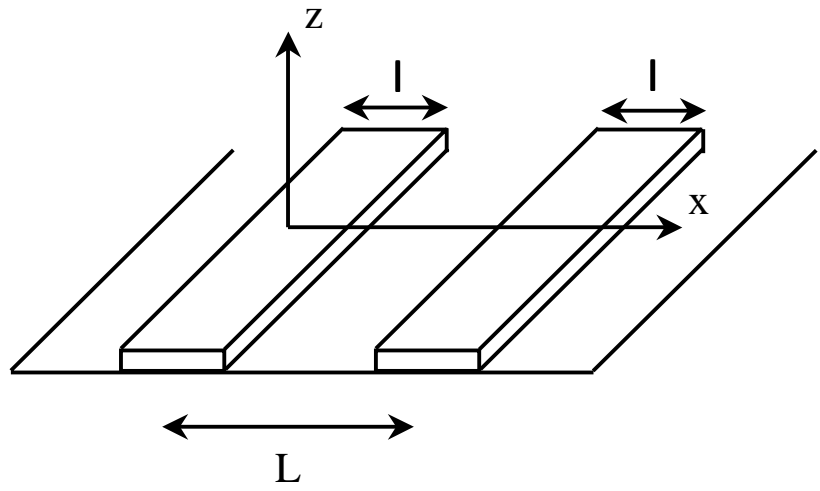

Figure 2 :

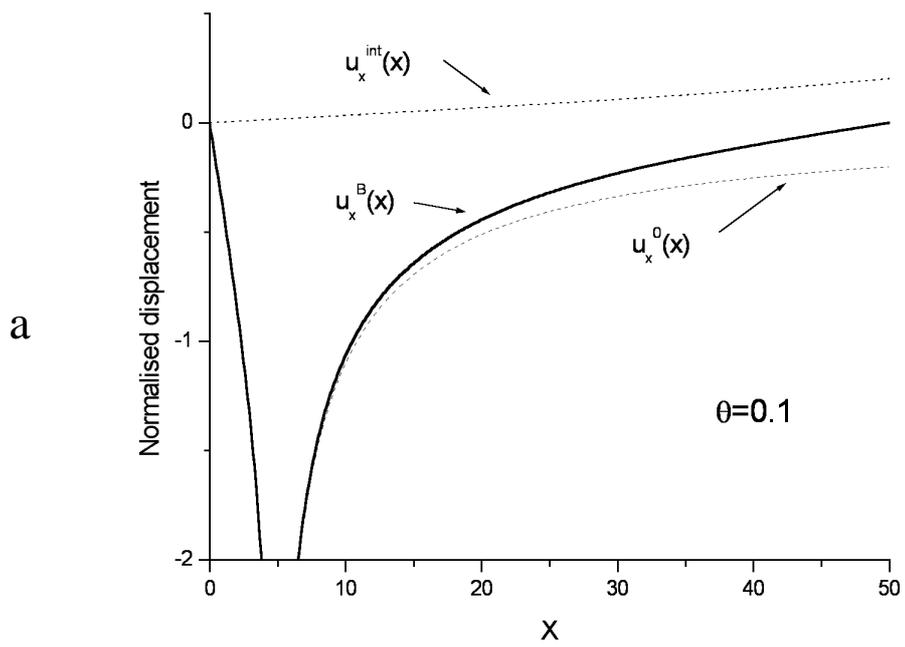

a

b

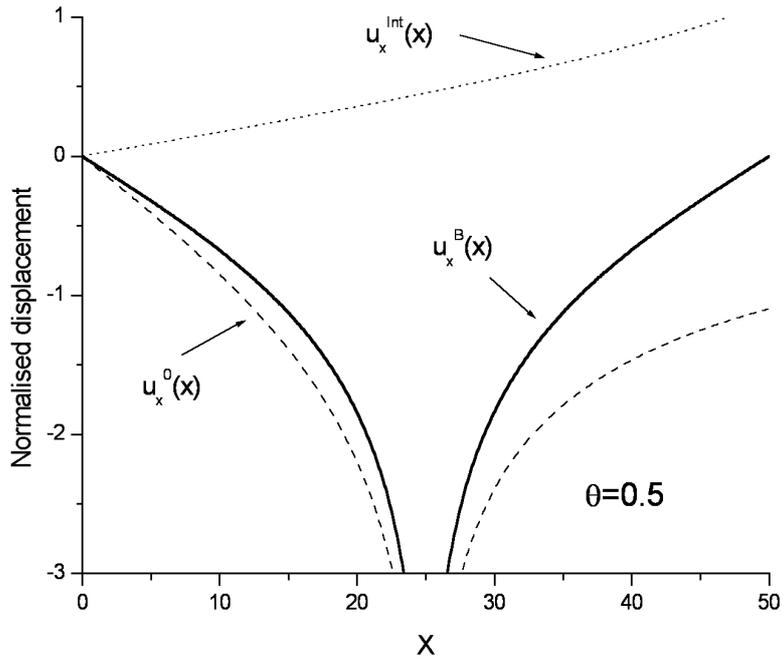

c

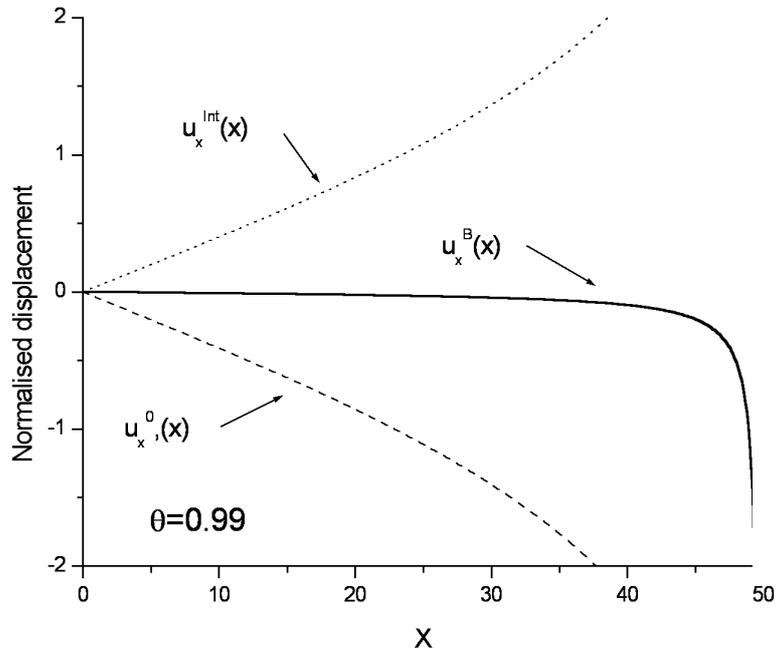

Fig 3

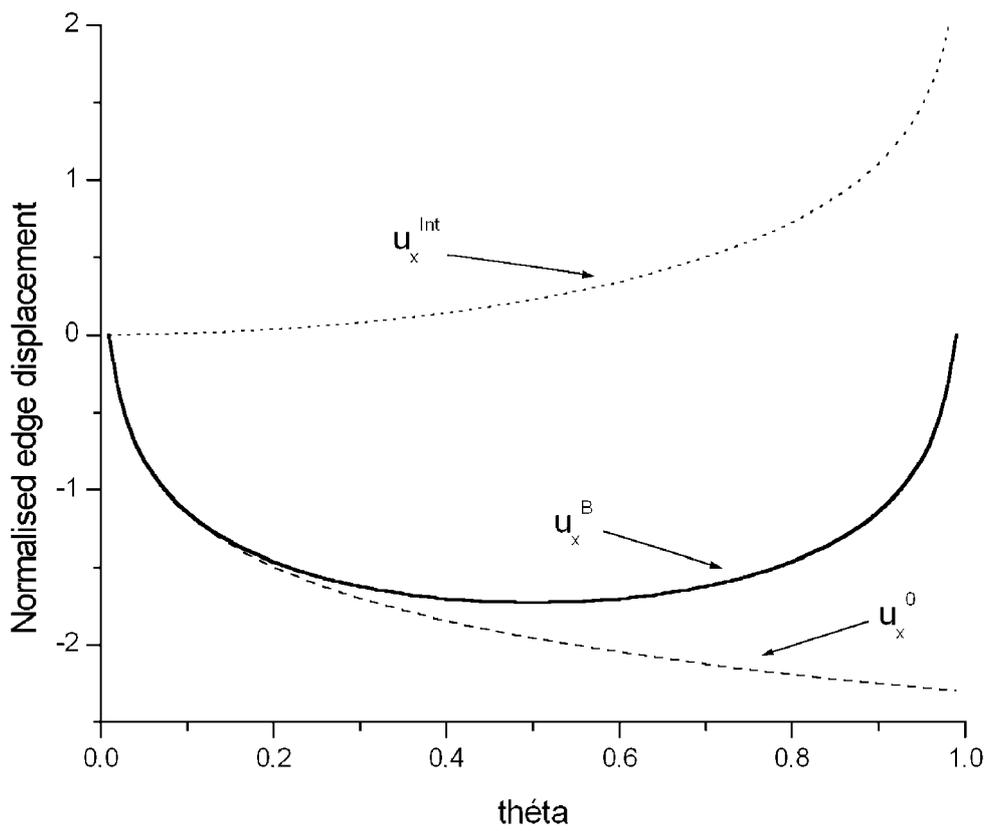

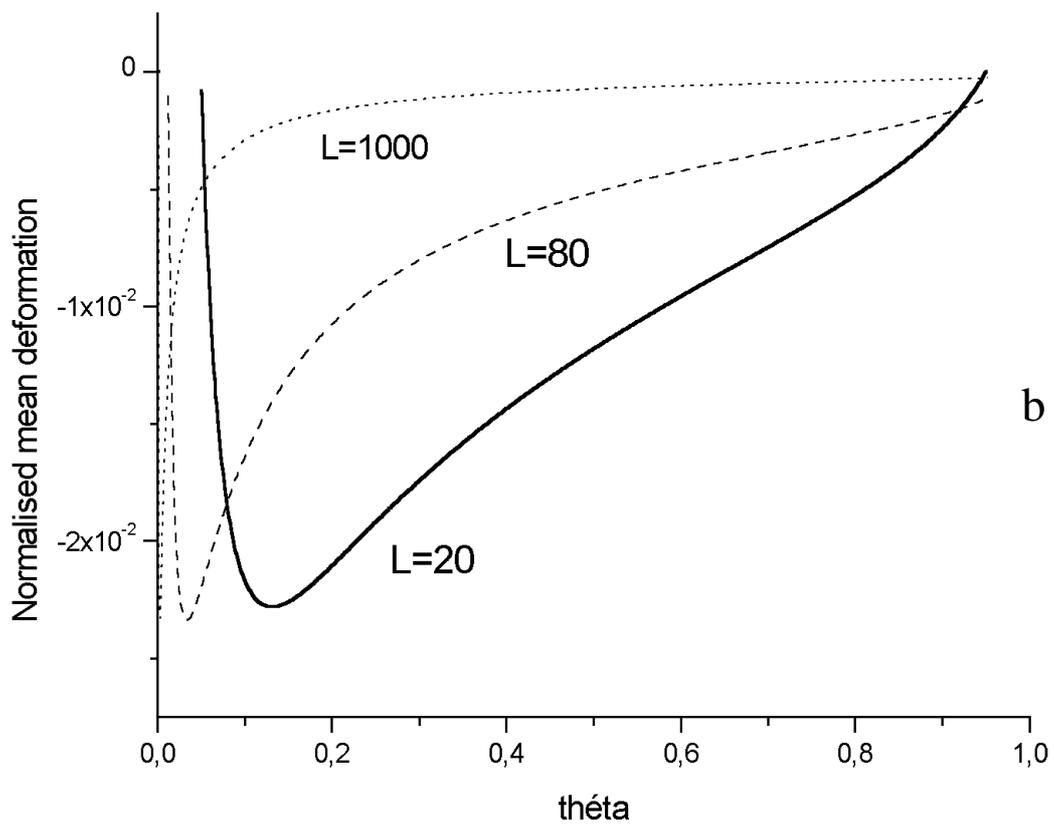

Fig4 :

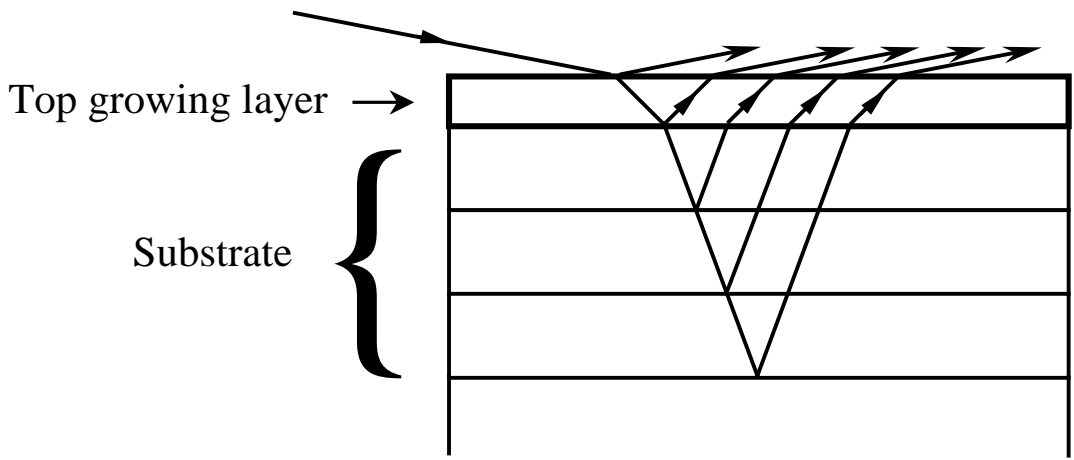

Figure 5 :

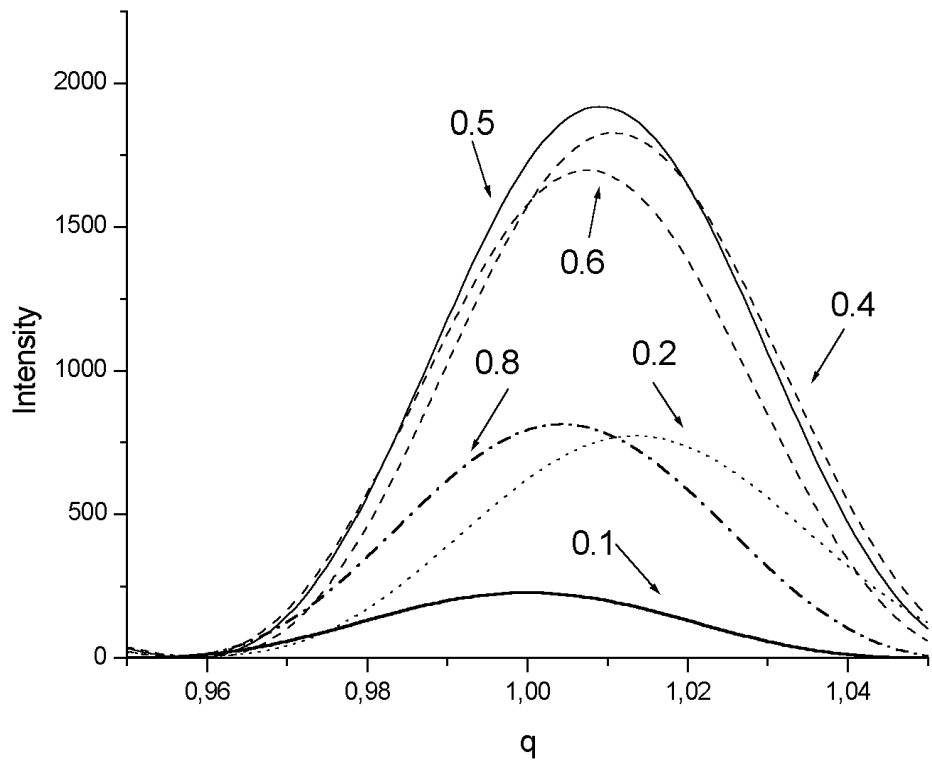

Figure 6 :

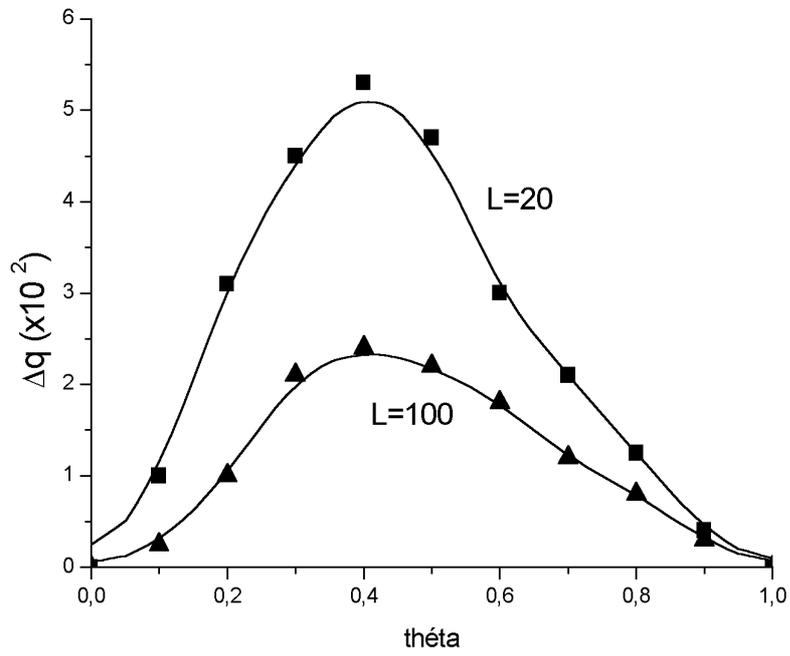

Figure 7 :

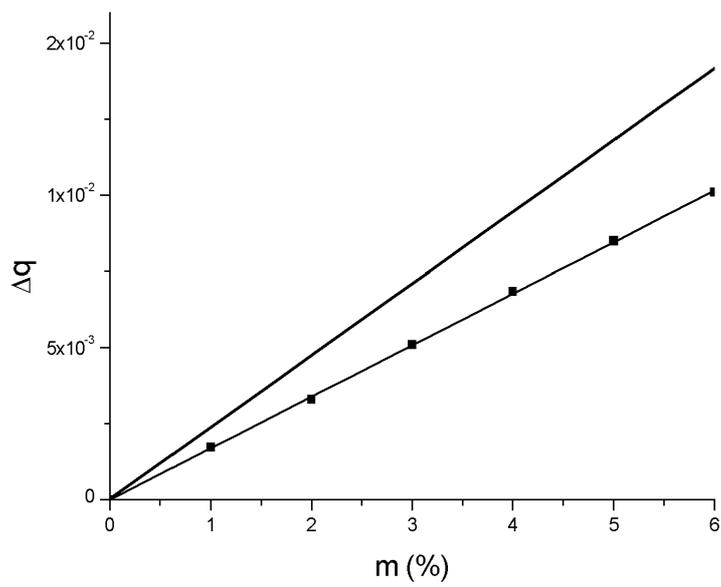

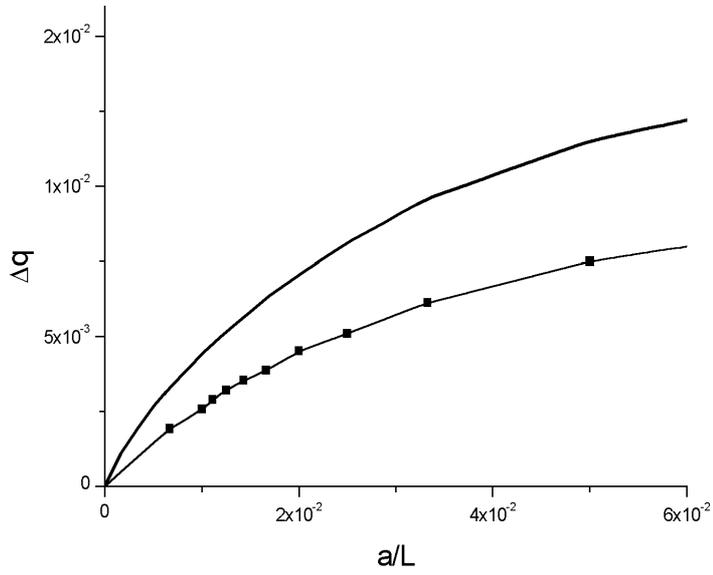

Figure 8 :

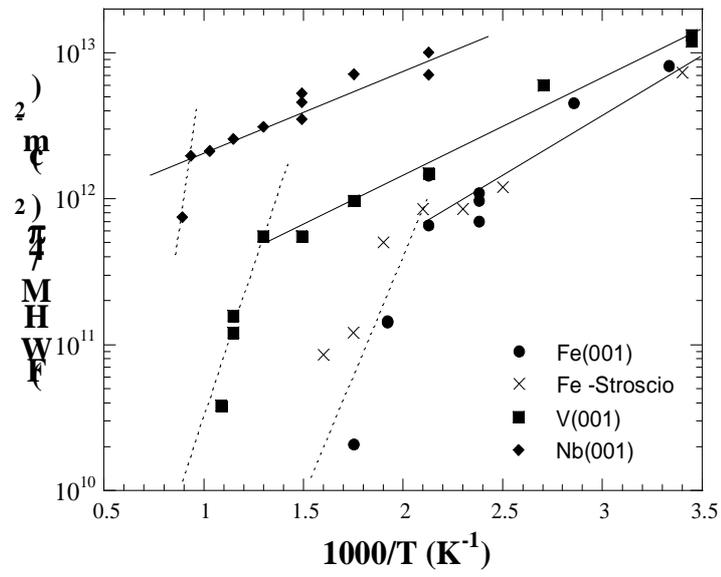

Figure 9

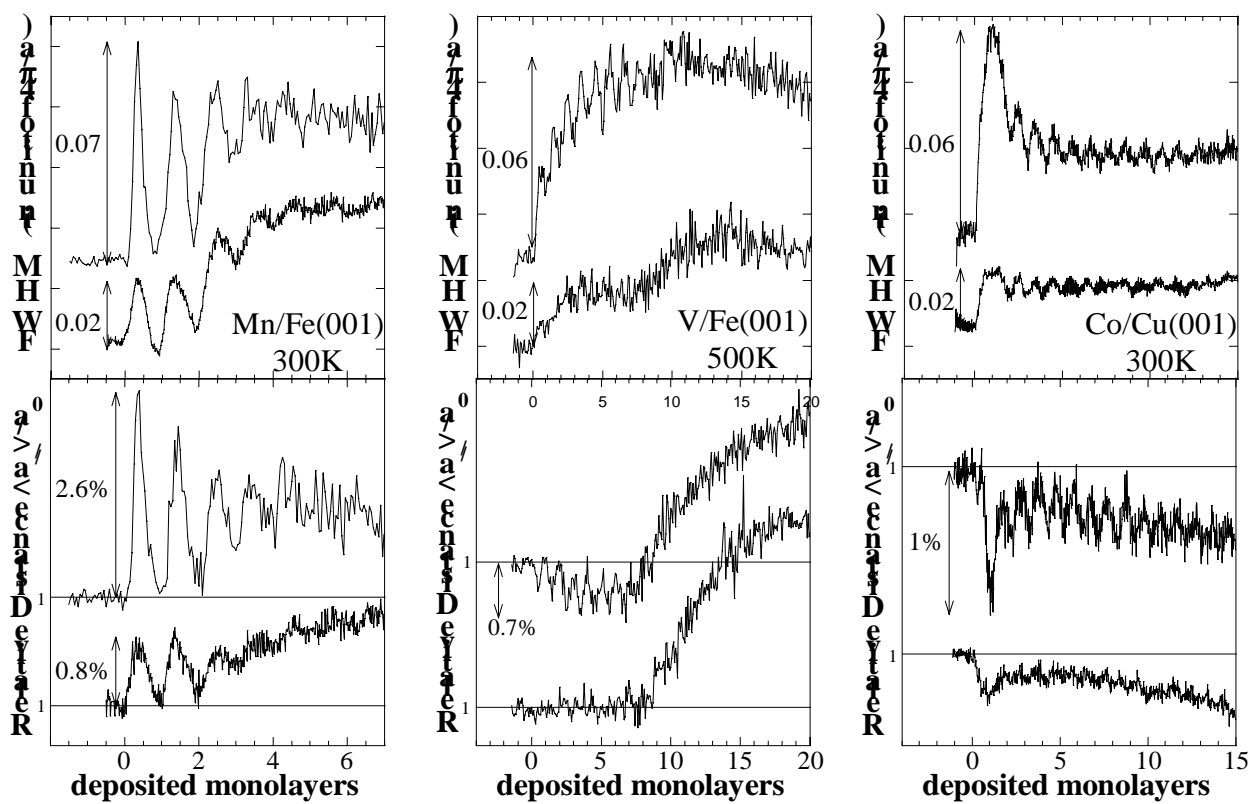

Figure 10 :

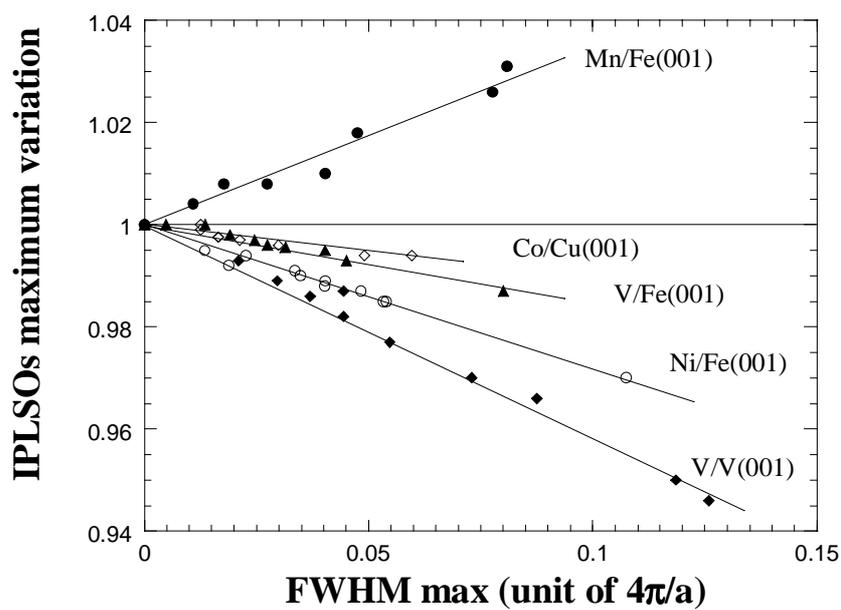